\documentclass{amsart}
\usepackage{amsthm}
\usepackage{graphicx,textcomp}
\usepackage{amsmath,bm,amsfonts,mathrsfs,amssymb}
\usepackage{hyperref}
\newtheorem{theorem}{Theorem}
\newtheorem{lemma}[theorem]{Lemma}
\newtheorem{prop}[theorem]{Proposition}

\newtheorem{rem}[theorem]{Remark}

\usepackage{appendix}
\usepackage[normalem]{ulem}
\usepackage{graphicx,color}
\usepackage{xcolor}

\counterwithout{equation}{section}   

\begin{document}
\title[Determinant of the Dirichlet-to-Neumann map]{The determinant of the Dirichlet-to-Neumann map for a surface with boundary and periods of holomorphic differentials on its double}

\author{Dmitrii Korikov}
	\address{St. Petersburg Department of Steklov Mathematical Institute
of Russian Academy of Sciences, 27 Fontanka, St. Petersburg, Russia, \url{https://orcid.org/0000-0002-3212-5874}}
	\email{dmitrii.v.korikov@gmail.com}
	
	\author{Alexey Kokotov}
	\address{Department of Mathematics \& Statistics, Concordia University, 1455 De Maisonneuve Blvd. W. Montreal, QC  H3G 1M8, Canada, \url{https://orcid.org/0000-0003-1940-0306}}
	\email{alexey.kokotov@concordia.ca}
	
\begin{abstract}
Let $(M,g)$ be a smooth orientable $2d$ Riemannian manifold of genus $\mathfrak{g}$ with  Riemannian metric $g$ and connected boundary $\Gamma$. Let $\Lambda$ be the Dirichlet-to-Neumann map on $\Gamma$ and let ${\rm det}_\zeta(\Lambda)$ be its (modified, i. e. with  zero mode excluded) $\zeta$-regularized determinant. It is well-known that the quantity ${\rm det}_\zeta(\Lambda)/|\Gamma|$ (where $|\Gamma|$ is the length of $\Gamma$) is a conformal invariant. It was shown by Edward and Wu (\cite{EV}) that this invariant equals one for $\mathfrak{g}=0$; in the case $\mathfrak{g}>0$ Guillarmou and Guillop\'e \cite{Guillarmou} found two explicit expressions for this invariant through the Rouelle and (respectively) the Selberg zeta-functions of the two surfaces of negative constant curvature from the conformal class of $(M,g)$: one is  of infinite volume and complete whereas another has geodesic boundary. 

We present an elementary counterpart of the formulae of Guillarmou and Guillop\'e using  the periods of holomorphic differentials on the double $2M$ of $M$ only.  Our approach is based on the properties of the Hilbert transform of $M$ \cite{B,HilbKor} and a version of the Friedlander-Guillemin regularization of the determinants of pseudodifferenial operators \cite{KV,Ww,F}. In particular, a connection between the length spectra of (uniformized) $M$, $2M$  and the periods of holomorphic differentials on $2M$  is established.
\end{abstract}

	\maketitle

\section*{Introduction}
Let $(M,g)$ be a (smooth) orientable surface of genus $\mathfrak{g}$ with metric $g$ and connected boundary $\Gamma$. We assume that $\Gamma$ is parametrized by the arc length and identify it with the boundary of the disk $\mathbb{D}$ of radius $|\Gamma|/2\pi$. The Dirichlet-to-Neumann (DN  in the sequel) map of $(M,g)$ is the self-adjoint operator in $L_2(\Gamma,dl_g;\mathbb{C})$ given by $\Lambda f:=\partial_\nu u^f|_\Gamma$, where $u^f$ is the harmonic extension of $f\in H^{1}(\Gamma;\mathbb{C})\equiv {\rm Dom}(\Lambda)$ into $(M,g)$ and $\nu$ is the unit exterior normal to $\Gamma$. Let $\partial_\gamma$ denotes differentiation with respect to the length along $\Gamma$ (in what follows, we choose the orientation on $M$ in such a way that $\gamma$ is obtained from the $\nu$ via the counter-clockwise rotation by the right angle). It is well-known \cite{LeeU} that $\Lambda$ is a $\Psi$DO of order one, admitting decomposition
\begin{equation}
\label{DN PDO decomp}
\Lambda=\Lambda_0+\dot{\Lambda}, \qquad \Lambda_0=|\partial_\gamma|,
\end{equation}
where $\dot{\Lambda}$ is smoothing $\Psi$DO. Note that $\Lambda_0$ is the DN map of $\mathbb{D}$.

In the sequel we restrict all the operators to the Hilbert space $\mathscr{H}:={\rm Ran}(\Lambda)=L_2(\Gamma,dl_g;\mathbb{C})\ominus\mathbb{C}$, thereby making $\Lambda$ and $\partial_\gamma$ invertible.

It is a common lore that the operator $\zeta$-function of the Dirichlet-to-Neumann operator (in any dimension) is regular at $s=0$, however, it is hard to find the precise reference for this result in the literature (cf. the Remark after Lemma 1.8.2 in \cite{Gilkey}). Presumably, it should follow from formula (4.2.62), \cite{Grubb} supplied with some special considerations ("no pure $\log$-term in the heat asymptotics"); we derive this property of $\zeta_{\Lambda}(s)$ independently in Proposition \ref{zeta analyt fin} below.
 
The (modified) $\zeta$-regularized determinant of $\Lambda$ is defined as ${\rm det}_\zeta(\Lambda):={\rm exp}\big(-\zeta'_\Lambda(0)\big)$; according to  \cite{Guillarmou} it is related to the length spectrum of $(M\backslash\Gamma,h_\infty)$ (where $h_\infty$ is the unique complete hyperbolic metric on $M\backslash\Gamma$) via
\begin{equation}
\label{GG formula}
{\rm det}_\zeta(\Lambda)/|\Gamma|=\Big((2\pi s)^{-2\mathfrak{g}}\mathscr{R}(s)\Big)\Big|_{s=0}/(1-2\mathfrak{g})
\end{equation}
for $\mathfrak{g}>1$, where $\mathscr{R}(s):=\prod_{\gamma}(1-e^{-s\,|\gamma|_h})$ is the Ruelle zeta-function, the product is over all primitive closed geodesics in $(2M,h_\infty)$ (in particular, the left-hand side of (\ref{GG formula}) is a conformal invariant). (There is a similar relation also established in  \cite{Guillarmou} with the Selberg zeta-function  of the hyperbolic surface with geodesic boundary conformally equivalent to $M$, we do not quote this result in full for brevity's sake.)

For the trivial case $\mathfrak{g}=0$, the explicit formulas $\zeta_{\Lambda_0}(s)=2\zeta(s)\Big(\frac{2\pi}{|\Gamma|}\Big)^{-s}$, $\zeta_{\partial_\gamma}(s)=\zeta_{\Lambda_0}(s){\rm cos}\Big(\frac{\pi s}{2}\Big)$ and $\zeta(0)=-\frac{1}{2}$, $\zeta'(0)=-\frac{{\rm log}(2\pi)}{2}$ (where $\zeta$ is the Riemann zeta function) imply
\begin{equation}
\label{GG formula disk}
{\rm det}_\zeta(\Lambda_0)={\rm det}_\zeta(\partial_\gamma)=|\Gamma|.
\end{equation}
(In \cite{EV} it is shown that  the same  expression holds true for ${\rm det}_\zeta(\Lambda)$ if $M$ is simply connected.)

Following Belishev\cite{B,BKChar}, introduce the {\it Hilbert transform} of $M$ by
\begin{equation*}
H:=\partial_\gamma^{-1}\Lambda.
\end{equation*}
It is easy to see that the operator $f+\mathbb{C}\mapsto Hf+\mathbb{C}$ acting on the quotient space $L_2(\Gamma,dl_g;\mathbb{C})/\mathbb{C}$ depends only on the complex structure on $M$ (if $H'$ is the Hilbert transfrom of $(M,h:=e^\phi g)$, then $H'f=Hf+c(f)$, where $c(f)$ is the constant providing the orthogonality of $H'f$ to the constants in $L_2(\Gamma,dl_h;\mathbb{C})$). Conversely, the Hilbert transform $H$ determines the surface $M$ up to biholomorphism \cite{B}. 

Decomposition (\ref{DN PDO decomp}) implies
\begin{equation}
\label{Hilbert PDO decomp}
H=H_0+\dot{H}, \qquad H_0=\partial_\gamma^{-1}|\partial_\gamma|,
\end{equation}
where $\dot{H}$ is a smoothing $\Psi$DO and $H_0$ is the Hilbert transform of $\mathbb{D}$. Note that $H_0^2=-I$. The spectrum of $H$ is described in \cite{HilbKor}. The essential spectrum of $H$ consists of two eigenvalues $-i,+i$; the corresponding eigenspaces ${\rm Ker}(H+i)$ (${\rm Ker}(H-i)$) are spanned by the boundary traces of holomorphic (anti-holomorphic) functions with square-integrable differentials on $M$. The discrete spectrum of $H$ consists of the eigenvalues (having the same algebraic and geometric multiplicities and counted with respect to them)
$$\lambda_{\pm 1}(H)=\pm i\mu_1,\dots,\lambda_{\pm\mathfrak{g}}(H)=\pm i\mu_{\mathfrak{g}}.$$
The numbers $0<\mu_1\le\mu_2\le\dots\le\mu_{\mathfrak{g}}<1$ are related to the periods of Abelian differentials on the Schottky double $2M$ of $M$. 
\begin{lemma}[see Corollary 4, \cite{znsbKor}] \label{Kolemma} {\rm 1.} Introduce the space $\mathscr{H}^1_0(M)$ of harmonic one-forms on $M$ vanishing along $\Gamma$. Let $\{l_k\}_{k=l}^{2\mathfrak{g}}$ be a basis in $H_1(M, \partial M; {\mathbb Z})$, and $\{\upsilon_k\}_{k=l}^{2\mathfrak{g}}$ be the dual basis in $\mathscr{H}^1_0(M)$ obeying $\int_{l_i}\upsilon_j=\delta_{ij}$. Then $\{\lambda_{\pm k}\}_{k=1}^{\mathfrak{g}}$ are the eigenvalues of the period matrix $\mathfrak{B}$ with entries $\mathfrak{B}_{ij}=\int_{l_i}\star\upsilon_j$.

{\rm 2.} Let $\tau$ be the antiholomorphic involution $\tau$, such that $2M/\tau\equiv M$. Let $\{a_j\}_{j=1}^{2\mathfrak{g}}$ be cycles that extend to a symplectic basis of $H_1(2M;\mathbb{Z})$ and obey the anti-symmetry condition $a_j=-\tau\circ a_j$. Then
$$2^{2\mathfrak{g}}{\rm det}(\Im\mathbb{B})={\rm det}(\mathfrak{B})=\prod_{k,\pm}\lambda_{\pm k}$$
is valid for the b-period matrix $\mathbb{B}$ of $2M$ corresponding to any canonical homology basis extending $\{a_j\}_{j=1}^{2\mathfrak{g}}$.
\end{lemma}
(For the convenience of the reader, the proof of this statement is presented in the appendix.)

Due to the presence of the essential spectrum $\{-i,+i\}$, neither the zeta-regularized nor the Fredholm determinant of $H$ can be defined. We simply define  the determinant of $H$ via
\begin{equation}
\label{toy det}
{\rm DET}(H):=\lambda_1\cdot\dots\cdot\lambda_{\mathfrak{g}}\cdot\lambda_{-1}\cdot\dots\cdot\lambda_{-\mathfrak{g}}={\rm det}(H_{\rm disc}).
\end{equation}
i. e., by restricting $H$ to its discrete spectrum subspace $\mathcal{H}_{\rm disc}:=\bigoplus\limits_{\pm k}{\rm Ker}(H-\lambda_k)$ of dimension $2\mathfrak{g}$; such a restriction is denoted by $H_{\rm disc}$ (thus, roughly speaking, we formally put $(-i)(+i)(-i)(+i)\dots=1$ for the eigenvalues of $H$ of infinite multiplicity, by this definition, we have ${\rm DET}(H_0)=1$).

\smallskip

In this note, we prove the following relation 
\begin{equation}
\label{dumb formula}
{\rm det}_\zeta(\Lambda)={\rm det}_\zeta(\partial_\gamma)\cdot{\rm DET}(H).
\end{equation}
As a corollary of (\ref{dumb formula}), (\ref{GG formula disk}), (\ref{toy det}), and (\ref{GG formula}), one arrives at
\begin{align*}
(\mu_1\cdot\dots\cdot\mu_{\mathfrak{g}})^2={\rm DET}(H)={\rm det}_\zeta(\Lambda)/{\rm det}_\zeta(\partial_\gamma)=\\
={\rm det}_\zeta(\Lambda)/|\Gamma|=\Big((2\pi s)^{-\mathfrak{g}}\mathscr{R}(s)\Big)\Big|_{s=0}/(1-2\mathfrak{g}).
\end{align*}
Thus, we have obtained the formula
\begin{equation}
\label{KoKoFormula}
\frac{\Big((2\pi s)^{-2\mathfrak{g}}\mathscr{R}(s)\Big)\Big|_{s=0}}{1-2\mathfrak{g}}=(\mu_1\cdot\dots\cdot\mu_{\mathfrak{g}})^2 ={\rm det}(\mathfrak{B})=2^{2\mathfrak{g}}{\rm det}(\Im\mathbb{B})
\end{equation}
that relates the length spectrum on $(M\backslash\Gamma,h_\infty)$ with the periods of certain basic Abelian differentials on the double $(2M,\tau)$ or the periods of harmonic differentials on $M$ vanishing along $\Gamma$.

\section*{Proof of formula (\ref{dumb formula})}

{\bf 1)} Introduce the family
\begin{align}
\label{Lambda familiy}
\Lambda_t:=t\Lambda+(1-t)\Lambda_0=\Lambda_0+t\dot{\Lambda}.
\end{align}
Since $\Lambda_t$ is a convex combination of two positive-definite operators $\Lambda,\Lambda_0$, it is also positive-definite and its spectrum is belongs to $[2r_0,+\infty)$, where $r_0>0$. The map $t,\lambda\mapsto (\lambda-\Lambda_t)^{-1}$ ($t\in[0,1]$, $\lambda\in\mathbb{C}\backslash[2r_0,+\infty)$ is continuous in the norm $\|\cdot\|:=\|\cdot\|_{B(\mathcal{H})}$ and obeys the (uniform in $t$) estimate
\begin{equation}
\label{estimate L2}
\|(\lambda-\Lambda_t)^{-1}\|=O(|\lambda|^{-1}) \qquad (\Re\lambda\le r_0).
\end{equation}
As a corollary of (\ref{estimate L2}), we have 
\begin{equation}
\label{estimate L2 1}
\|\Lambda_t(\Lambda_t-\lambda)^{-1}\|=\|\lambda(\Lambda_t-\lambda)^{-1}-I\|=O(1) \qquad (\Re\lambda\le r_0).
\end{equation}

Introduce the Sobolev norms
$$\|f\|_{H^s(\Gamma)}:=\Big(\sum_{k\in\mathbb{Z}}|\hat{f}_k|^2(1+|k|^2)^s\Big)^{1/2},$$
where $\hat{f}_k$ are the coefficients of the Fourier expansion of $f$ on $\Gamma$. Since $\Lambda_0=|\partial_\gamma|$, we have $(\widehat{\Lambda f})_k=|k|\hat{f}_k$. Thus, for each $C>0$, the Sobolev norm $\|f\|_{H^{1}(\Gamma)}$ is equivalent to $\|\Lambda_0 f\|_{L_2(\Gamma)}+C\|f\|_{L_2(\Gamma)}$. Then formula (\ref{Lambda familiy}) implies
\begin{align*}
\|f\|_{H^{1}(\Gamma)}\le c\big(\|\Lambda_t f\|_{L_2(\Gamma)}+(C+t\|\dot{\Lambda}\|_{B(L_2(\Gamma))})\|f\|_{L_2(\Gamma)}\big).
\end{align*} 
Replacing $f$ by $(\Lambda_t-\lambda)^{-1}f$ in the last inequality and taking into account (\ref{estimate L2}), (\ref{estimate L2 1}), one obtains
\begin{equation}
\label{Sobolev boundedness}
\|(\Lambda_t-\lambda)^{-1}\|_{B(\mathscr{H};H^1(\Gamma))}=O(1) \quad (\Re\lambda\le r_0).
\end{equation}

In view of the definition of Sobolev norms and the H\"older inequality 
$$\sum_{k\in\mathbb{Z}}|\hat{f}_k|^2(1+|k|^2)^{q}\le \Big(\sum_{k\in\mathbb{Z}}|\hat{f}_k|^2\Big)^{1-q}\Big(\sum_{k\in\mathbb{Z}}|\hat{f}_k|^2(1+|k|^2)\Big)^{q},$$
the following interpolation inequality
$$\|u\|_{H^{q}(\Gamma)}\le C_q\|u\|_{H^{1}(\Gamma)}^{q}\|u\|_{L_2(\Gamma)}^{1-q} \qquad (u\in H^{1}(\Gamma))$$ 
is valid for $q\in (0,1)$. Substituting $u=(\Lambda_t-\lambda)^{-1}f$ into this inequality and taking into account (\ref{Sobolev boundedness}), one arrives at
\begin{equation}
\label{Sobolev half decay}
\|(\Lambda_t-\lambda)^{-1}\|_{B(\mathscr{H};H^{q}(\Gamma))}=O(|\lambda|^{q-1})  \quad (\Re\lambda\le r_0)
\end{equation}

{\bf 2)} The standard formula of holomorphic calculus reads
$$\log \Lambda_t=\frac{1}{2\pi i}\int\limits_{\mathfrak{C}}\Lambda_t(\lambda-\Lambda_t)^{-1}\frac{{\rm log}(\lambda)}{\lambda}d\lambda,$$
where $\mathfrak{C}(r)$ is a vertical line $y\mapsto\lambda(y)=r+iy$, $0<r<r_0$, and the cut of the logarithm is $(-\infty,0]$. Then
$$\partial_{t}\Big[\Lambda^{-s}{\rm log}(\Lambda_t)\Big]=\frac{1}{2\pi i}\int\limits_{\mathfrak{C}}\Lambda^{-s}\partial_t\Big[\Lambda_t(\lambda-\Lambda_t)^{-1}\Big]\frac{{\rm log}(\lambda)}{\lambda}d\lambda=\mathfrak{J}_1+\mathfrak{J}_2,$$
where
\begin{align*}
\mathfrak{J}_1:&=\frac{1}{2\pi i}\int\limits_{\mathfrak{C}}\Lambda^{-s}\dot{\Lambda}(\lambda-\Lambda_t)^{-1}\frac{{\rm log}(\lambda)}{\lambda}d\lambda,\\
\mathfrak{J}_2:&=\frac{1}{2\pi i}\int\limits_{\mathfrak{C}}\Lambda^{-s}(\lambda-\Lambda_t)^{-1}\Lambda_t\dot{\Lambda}(\lambda-\Lambda_t)^{-1}\frac{{\rm log}(\lambda)}{\lambda}d\lambda.
\end{align*}

For any $s\in\mathbb{C}$, the operator $\Lambda^{-s}\dot{\Lambda}$ is a smoothing PDO (thus, it is of trace class), while  the map $t,\lambda\mapsto (\lambda-\Lambda_t)^{-1}$ is continuous in the norm $\|\cdot\|$ and obeys (\ref{estimate L2}). Due to the well-known inequality 
\begin{equation}
\label{trace norm of product}
\|AB\|_1=\|BA\|_1\le\|A\|_1\|B\|,
\end{equation}
where $\|\cdot\|_1$ is the trace norm, the integrand in $\mathfrak{J}_1$ is continuous in $t,\lambda$ in the trace norm, and its trace norm is $O(\lambda^{-2}{\rm log}(\lambda))$ uniformly in $t$. Thus, for any $s\in\mathbb{C}$, the integral $\mathfrak{J}_1$ converges in the trace norm. 

Next, the map $t\mapsto\Lambda_t\dot{\Lambda}=\Lambda_0\dot{\Lambda}+t\dot{\Lambda}^2$ is continuous in the trace norm. Due to estimate (\ref{Sobolev half decay}), we have 
\begin{equation}
\label{lambda -s resolvent est}
\|\Lambda^{-s}(\lambda-\Lambda_t)^{-1}\|=O(|\lambda|^{q-1}) \qquad (\Re s\ge -q),
\end{equation}
where $q\in (0,1)$ is arbitrary. In particular, (\ref{lambda -s resolvent est}) and formula $\partial_\lambda(\Lambda^{-s}(\lambda-\Lambda_t)^{-1})=-\Lambda^{-s}(\Lambda^{-s}(\lambda-\Lambda_t)^{-2})$ imply that the map $\lambda\mapsto\Lambda^{-s}(\lambda-\Lambda_t)^{-1}$ ($\lambda\in\mathbb{C}\backslash[2r_0,+\infty)$) is continuous in the norm $\|\cdot\|$ for $s>-1$. Due to these facts and (\ref{trace norm of product}), the integrand in $\mathfrak{J}_2$ is continuous in $t,\lambda$ in the trace norm, and its trace norm is $O(|\lambda^{q-3}{\rm log}\lambda|)$ for $\Re s>-q$. Hence, the integral $\mathfrak{J}_2$ converges in the trace norm for $\Re s>-1$.

Similarly, we have
\begin{align*}
\partial_\sigma\mathfrak{J}_1:&=\frac{\partial }{\partial\sigma}\,\frac{1}{2\pi i}\int\limits_{\mathfrak{C}}\Lambda^{-s}\,{\rm log}\Lambda\,\dot{\Lambda}(\lambda-\Lambda_t)^{-1}\frac{{\rm log}(\lambda)}{\lambda}d\lambda,\\
\partial_\sigma\mathfrak{J}_2:&=\frac{\partial }{\partial\sigma}\,\frac{1}{2\pi i}\int\limits_{\mathfrak{C}}\Lambda^{-s}\,{\rm log}\Lambda\,(\lambda-\Lambda_t)^{-1}\Lambda_t\dot{\Lambda}(\lambda-\Lambda_t)^{-1}\frac{{\rm log}(\lambda)}{\lambda}d\lambda.
\end{align*}
where $\sigma=\Re s$ or $\sigma=\Im s$. Let us represent $\Lambda^{-s}\,{\rm log}\Lambda\,(\lambda-\Lambda_t)^{-1}$ as product of two operators $[\Lambda^{-s-q}\,{\rm log}\Lambda]$ and $(\Lambda^q(\lambda-\Lambda_t)^{-1})$. We have
$$\Lambda^{-q-s}\log \Lambda=\frac{1}{2\pi i}\int\limits_{\mathfrak{C}}\Lambda^{1-q-s}(\lambda-\Lambda)^{-1}\frac{{\rm log}(\lambda)}{\lambda}d\lambda,$$
where the norm of the integrand is $O(|\lambda^{-q-\Re s-1}{\rm log}\lambda|)$ due to inequality (\ref{lambda -s resolvent est}) (with $s$ replaced by $1-q-s$). Thus, $\Lambda^{-q-s}\log \Lambda\in B(H)$ for $\Re s>-q$. Estimating the second operator by the use of (\ref{lambda -s resolvent est}), one arrives at $\|\Lambda^{-s}\,{\rm log}\Lambda\,(\lambda-\Lambda_t)^{-1}\|=O(|\lambda^{q-1})$ for $\Re s>-q$. Now repeating the arguments from the proof of convergence of $\mathfrak{J}_1,\mathfrak{J}_2$ we obtain that $\mathfrak{J}_1,\mathfrak{J}_2$ are differentiable in $\Re s$,$\Im s$ with respect to the trace norm for $\Re s>-1$. Since $\partial_{\overline{s}}\Lambda^{-s}=0$, we have proved that $\mathfrak{J}_1,\mathfrak{J}_2$ are holomorphic in $s$ for $\Re s>-1$. This yields the following statement
\begin{prop}
\label{zeta analyt 1}
The function $s\mapsto\partial_{t}{\rm Tr}\Big[\Lambda^{-s}{\rm log}(\Lambda_t)\Big]$ is well-defined and holomorphic for $\Re s>-1$. 
\end{prop}
Since the trace of the product of operators does not change under their cyclic permutations, we have
\begin{align*}
\partial_{t}{\rm Tr}\Big[\Lambda^{-s}{\rm log}(\Lambda_t)\Big]\Big|_{s=0}=\frac{1}{2\pi i}\int\limits_{\mathfrak{C}}{\rm Tr}\Big((\lambda-\Lambda_t)^{-1}\dot{\Lambda}+(\lambda-\Lambda_t)^{-2}\Lambda_t\dot{\Lambda}\Big)\frac{{\rm log}(\lambda)}{\lambda}d\lambda.
\end{align*}
In view of the equality $-\lambda\partial_\lambda(\lambda-\Lambda_t)^{-1}=(\lambda-\Lambda_t)^{-1}+(\lambda-\Lambda_t)^{-2}\Lambda_t$ and formula (\ref{estimate L2}), the integration by parts yields
\begin{align}
\label{FG det variation}
\begin{split}
\partial_{t}{\rm Tr}\Big[\Lambda^{-s}{\rm log}(\Lambda_t)\Big]\Big|_{s=0}=\frac{-1}{2\pi i}\int\limits_{\mathfrak{C}}{\rm Tr}\Big(\partial_\lambda(\lambda-\Lambda_t)^{-1}\dot{\Lambda}\Big){\rm log}(\lambda)d\lambda=\\
=\frac{1}{2\pi i}\int\limits_{\mathfrak{C}}{\rm Tr}\Big((\lambda-\Lambda_t)^{-1}\dot{\Lambda}\Big)\frac{d\lambda}{\lambda}={\rm Tr}(\Lambda_t^{-1}\dot{\Lambda}).
\end{split}
\end{align}

{\bf 3)} Since
$$\Lambda_t^{-s}=\frac{1}{2\pi i}\int_{\mathfrak{C}(r)}(\lambda-\Lambda_t)^{-1}\lambda^{-s}d\lambda,$$
we have
\begin{equation}
\label{derivative reqularizer}
\partial_t[\Lambda_t^{-s}{\rm log}(\Lambda_0)]=\frac{1}{2\pi i}\int_{\mathfrak{C}(r)}\Big[(\lambda-\Lambda_t)^{-1}\dot{\Lambda}(\lambda-\Lambda_t)^{-1}{\rm log}(\Lambda_0)\Big]\lambda^{-s}d\lambda.
\end{equation}
Let us estimate the trace norm of the integrand in the right-hand side of (\ref{derivative reqularizer}). First, the equality ${\rm log}(\Lambda_0)(e^{\pm \frac{ikx}{|\Gamma|}})={\rm log}|k|\,e^{\pm \frac{ikx}{|\Gamma|}}$ (where $k\in\mathbb{N}$ and $x$ is the natural parameter on $|\Gamma|$) means that ${\rm log}(\Lambda_0)\in B(H^{1/2}(\Gamma),L_2(\Gamma))$. Denote $A=(\Lambda_t-\lambda)^{-1}{\rm log}(\Lambda_0)$, then $A^*={\rm log}(\Lambda_0)(\Lambda_t-\overline{\lambda})^{-1}$ and $\|A\|=\|A^*\|$ (if $\|A^*\|<\infty$, then $A$ extends to the bounded operator from the dense set $C^{\infty}(\Gamma)$ in $\mathscr{H}$). Therefore,
\begin{align}
\label{third term est}
\begin{split}
\|(\Lambda_t&-\lambda)^{-1}{\rm log}(\Lambda_0)\|=\|{\rm log}(\Lambda_0)(\Lambda_t-\overline{\lambda})^{-1}\|\le \\
\le& \|{\rm log}(\Lambda_0)\|_{B(H^{1/2}(\Gamma),L_2(\Gamma))}\|(\Lambda_t-\overline{\lambda})^{-1}\|_{B(\mathscr{H};H^{1/2}(\Gamma))}=O(|\lambda|^{-1/2})
\end{split}
\end{align}
for $\Re\lambda\le r_0$ due to (\ref{Sobolev half decay}).

Since $\dot{\Lambda}$ is a smoothing $\Psi$DO, it is of trace class and applying formula (\ref{trace norm of product}) twice yields
\begin{equation}
\label{integrand est trace norm}
\begin{split}
\|(\lambda-\Lambda_t)^{-1}\dot{\Lambda}(\lambda-\Lambda_t)^{-1}{\rm log}(\Lambda_0)\|_1&\le \\ 
\le \|(\Lambda_t-\lambda)^{-1}\|\|\dot{\Lambda}\|_1&\|(\Lambda_t-\lambda)^{-1}{\rm log}(\Lambda_0)\|=O(|\lambda|^{-3/2})
\end{split}
\end{equation}
due to (\ref{estimate L2}), (\ref{third term est}). As a corollary of (\ref{integrand est trace norm}), the right-hand side of (\ref{derivative reqularizer}) is well-defined and continuous in trace norm for $\Re s>-1/2$. Thus, we  get the following statement.
\begin{prop}
\label{zeta analyt 2}
The function $s\mapsto\partial_t[\Lambda_t^{-s}{\rm log}(\Lambda_0)]$ is well-defined and holomprphic for $\Re s>-1/2$. 
\end{prop}
Taking $s=0$ in (\ref{derivative reqularizer}), we obtain
\begin{align*}
\partial_t[\Lambda_t^{-s}{\rm log}(\Lambda_0)]\Big|_{s=0}=\frac{1}{2\pi i}\int_{\mathfrak{C}(r)}\Big[(\lambda-\Lambda_t)^{-1}\dot{\Lambda}(\lambda-\Lambda_t)^{-1}{\rm log}(\Lambda_0)\Big]\,d\lambda.
\end{align*}
Since the multivalued functions disappear in the right-hand side, one can shift the contour by decreasing $r$. Then estimate (\ref{integrand est trace norm}) leads to
\begin{align*}
\|\partial_t[\Lambda_t^{-s}{\rm log}(\Lambda_0)]\Big|_{s=0}\|_1\le c\int_{\mathfrak{C}(r)}|\lambda|^{-3/2}d\lambda=O(|r|^{-1/2}) \qquad (r\to -\infty).
\end{align*}
Since $r<r_0$ is arbitrary, we obtain
\begin{equation}
\label{FG det variation regularizer}
\partial_t{\rm Tr}(\Lambda_t^{-s}{\rm log}(\Lambda_0))|_{s=0}=0.
\end{equation} 

{\bf 4)} For large positive $\Re s$, we have
\begin{align}
\label{integrated fomula}
\begin{split}
{\rm Tr}\Big(\Lambda^{-s}{\rm log}(\Lambda_t)\Big)&={\rm Tr}\Big(\Lambda^{-s}{\rm log}(\Lambda_0)\Big)+\int\limits_{0}^t\partial_\tau{\rm Tr}\Big(\Lambda^{-s}{\rm log}(\Lambda_\tau)\Big)d\tau=\\
={\rm Tr}\Big(\Lambda_0^{-s}{\rm log}(\Lambda_0)\Big)&+\int\limits_{0}^1\partial_\tau {\rm Tr}\Big([\Lambda_\tau^{-s}{\rm log}(\Lambda_0)]\Big)d\tau+\int\limits_{0}^t\partial_\tau{\rm Tr}\Big(\Lambda^{-s}{\rm log}(\Lambda_\tau)\Big)d\tau
\end{split}
\end{align}
As a corollary of formula (\ref{integrated fomula}), Propositions \ref{zeta analyt 1}, \ref{zeta analyt 2}, and the equalities
\begin{align*}
{\rm Tr}\Big(\Lambda_0^{-s}{\rm log}(\Lambda_0)\Big)=-\partial_s{\rm Tr}(\Lambda_0^{-s})=-\partial_s\zeta_{\Lambda_0}(s)=-2\zeta'(s),\\
{\rm Tr}\Big(\Lambda^{-s}{\rm log}(\Lambda)\Big)=-\partial_s{\rm Tr}(\Lambda^{-s})=-\partial_s\zeta_{\Lambda_0}(s),
\end{align*}
we get the following proposition.
\begin{prop}
\label{zeta analyt fin}
For any $t\in[0,1]$, the function $s\mapsto{\rm Tr}\Big(\Lambda^{-s}{\rm log}(\Lambda_t)\Big)$ admits meromorphic continuation to the half-plane $\Re s>-1/2$ with unique pole $s=1$ of order one. In particular, the zeta function of the DN map $\Lambda$ is meromorphic for $\Re s>-1/2$ and has the unique and simple pole $s=1$.
\end{prop}

Substituting $s=0$ into (\ref{integrated fomula}) and taking into account (\ref{FG det variation}) and (\ref{FG det variation regularizer}), we obtain
\begin{align*}
{\rm log\,det}(\Lambda)=-\partial_s\zeta_{\Lambda}(0)=a.c.\big[{\rm Tr}\Big(\Lambda^{-s}{\rm log}(\Lambda_1)\Big)\big]\Big|_{s=0}=\\
={\rm Tr}\Big(\Lambda_0^{-s}{\rm log}(\Lambda_0)\Big)+0+\int_{0}^1\partial_t{\rm Tr}\Big(\Lambda^{-s}{\rm log}(\Lambda_t)\Big)dt=\\
={\rm log\,det}(\Lambda_0)+\int_{0}^1{\rm Tr}(\Lambda_t^{-1}\dot{\Lambda})dt
\end{align*}
(here $a.c.$ means the analytic continuation). Now, introduce the family
$$H_t=\partial_\gamma^{-1}\Lambda_t=H_0+t\dot{H},$$
where $t\in[0,1]$ and $\dot{H}=\partial_\gamma^{-1}\dot{\Lambda}$ is a smoothing PDO. Then we have $\Lambda_t^{-1}\dot{\Lambda}=(\partial_\gamma H_t)^{-1}\partial_\gamma\dot{H}=H_t^{-1}\dot{H}$ and the last formula can be rewritten as 
\begin{align}
\label{anomaly deriving 1}
{\rm log}\Big(\frac{{\rm det}(\Lambda)}{{\rm det}(\Lambda_0)}\Big)=\int\limits_{0}^1{\rm Tr}(H_t^{-1}\dot{H})dt.
\end{align}

{\bf 5)} Introduce the family 
$$J_t:=-H_t^2=-(H_0+t\dot{H})^2=H_0^2+\tilde{H}_t=I+\tilde{H}_t$$
where $t\in[0,1]$. The operators $\tilde{H}_t:=-t(H_0\dot{H}+\dot{H}H_0)-t^2\dot{H}$ and $\partial_t J_t=-H_t\dot{H}+\dot{H}H_t$ are smoothing $\Psi$DOs and thus the essential spectrum of $J_t$ coincides with that of $H_0^2=-I$. Obviously, $\|J_t\|$ is bounded uniformly in $t\in[0,1]$.

If $H_t f=\lambda f$ ($f\ne 0$, $\lambda\ne\pm i$), then 
$$i\lambda(-i\partial_\gamma f,f)_{\mathscr{H}}=(\Lambda_t f,f)_{\mathscr{H}}\ge c\|f\|_{\mathscr{H}}^2,$$ 
where $c>0$ is independent of $t$. Since $-i\partial_\gamma$ is self-adjoint, $\lambda$ is pure imaginary and nonzero. In addition, since $(\lambda^2+1)\partial_\gamma f=\partial_\gamma(H_t^2+I)f=-\partial_\gamma\tilde{H}_t f$ and $\|\tilde{H}_t\|_{B(\mathscr{H};H^1(\Gamma))}$ is bounded uniformly in $t$, we have $|(-i\partial_\gamma f,f)_{\mathscr{H}}|\le c\|f\|^2_{\mathscr{H}}/(\lambda^2+1)$. This means that $(\lambda^2+1)\Im\lambda\ge c>0$ i.e. the spectrum of $H_t$ does intersect a fixed neighborhood of $\lambda=0$ for all $t\in[0,1]$. Hence, the following formula is valid
$${\rm log}(J_t)=\frac{1}{2\pi i}\int\limits_{\mathfrak{Q}}(\lambda-J_t)^{-1}{\rm log}(\lambda)d\lambda,$$
where $\mathfrak{Q}$ encloses the segment $[c_1,c_2]$ ($c_1,c_2>0$) containing the spectra of all $J_t=-H_t^2$ ($t\in[0,1]$). Differentiation in $t$ yields
$${\rm Tr}(\partial_t{\rm log}(J_t))=\frac{1}{2\pi i}\int\limits_{\mathfrak{Q}}{\rm Tr}[(\lambda-J_t)^{-1}\dot{J}_t(\lambda-J_t)^{-1}]{\rm log}(\lambda)d\lambda.$$
The operator $(\lambda-J_t)^{-1}$ is continuous in $t$ in the norm $\|\cdot\|$ while $\dot{J}_t$ is continuous in $t$ in the trace norm. Hence,
\begin{align*}
{\rm Tr}(\partial_t{\rm log}(J_t))=&\frac{1}{2\pi i}\int\limits_{\mathfrak{Q}}{\rm Tr}[(\lambda-J_t)^{-2}\dot{J}_t]{\rm log}(\lambda)d\lambda=\\
=&-{\rm Tr}\Bigg(\frac{1}{2\pi i}\int\limits_{\mathfrak{Q}}\partial_\lambda(\lambda-J_t)^{-1}\cdot{\rm log}(\lambda)d\lambda\dot{J}_t\Bigg)=\\
=&{\rm Tr}\Bigg(\frac{1}{2\pi i}\int\limits_{\mathfrak{Q}}(\lambda-J_t)^{-1}\frac{d\lambda}{\lambda}\dot{J}_t\Bigg)={\rm Tr}(J_t^{-1}\dot{J}_t)=\\
&={\rm Tr}(H_t^{-2}(H_t\dot{H}+\dot{H}H_t))=2{\rm Tr}(H_t^{-1}\dot{H}).
\end{align*}
Denote $K=H^2+I$, then $K$ is smoothing PDO whose essential spectrum is $\{0\}$ and the discrete spectrum consists of the eigenvalues $1-\mu_k^2\in(0,1)$ ($k=1,\dots,\mathfrak{g}$). In particular, $\|K\|<1$. Denote by $K_{\rm disc}$ the restriction of $K$ onto the ($2\mathfrak{g}$-dimensional) orthogonal complement to ${\rm Ker}(K)$ in $\mathscr{H}$ and note that $K_{\rm disc}=H^2_{\rm disc}+I$. Now, relation (\ref{anomaly deriving 1}) takes the form
\begin{align*}
{\rm log}\Big(\frac{{\rm det}(\Lambda)}{{\rm det}(\Lambda_0)}\Big)=\frac{1}{2}\int\limits_{0}^1{\rm Tr}(\partial_t{\rm log}(J_t))dt=\frac{1}{2}{\rm Tr}\Big(\int_{0}^1(\partial_t{\rm log}(J_t)dt\Big)=\\
=\frac{1}{2}{\rm Tr}({\rm log}(J_1)-{\rm log}(J_0))=\frac{1}{2}{\rm Tr}({\rm log}(-H^2))-0=\frac{1}{2}{\rm Tr}\,{\rm log}(I-K)=\\
=\frac{1}{2}\sum_{k=1}^{\infty}\frac{{\rm Tr}\big(K^k\big)}{k}=\frac{1}{2}\sum_{k=1}^{\infty}\frac{{\rm Tr}\big(K_{\rm disc}^k\big)}{k}=\frac{1}{2}{\rm Tr}\big({\rm log}(-H_{\rm disc}^2)\big)=\\
=\frac{1}{2}(-1)^{{\rm dim}(\mathcal{H}_{\rm disc})}{\rm log}({\rm det}(H_{\rm disc})^2)={\rm log}   ({\rm DET}(H)).
\end{align*}
So, we have obtained
$${\rm det}(\Lambda)={\rm det}(\Lambda_0)\,{\rm DET}(H)={\rm det}(\partial_\gamma)\,{\rm DET}(H).$$
By this, formula (\ref{dumb formula}) is proved. \qed

\section*{Appendix: proof of Lemma \ref{Kolemma}}

 1) Note that each $\upsilon\in\mathscr{H}^1_0$ admits the harmonic extension onto the double $2M$ that is anti-symmetric with respect to the involution, $\tau^*\upsilon=-\upsilon$. In particular, one has ${\rm dim}(\mathscr{H}^1_0)={\rm dim}(H^1(2M;K))=2 \mathfrak{g}$ and elements $\upsilon$ of $\mathscr{H}^1_0(M)$ are determined by their periods $c_{k}=\int_{l_k}\upsilon$ or its boundary values $\upsilon(\nu)$. 
 
Suppose that $\mathfrak{B}\vec{c}=\lambda\vec{c}$ and put $\upsilon=\sum_{k=1}^{2\mathfrak{g}}c_k\upsilon_k$. Then $\star\upsilon-\lambda\upsilon$ has zero periods on $M$ and thus it is exact, i.e. $\star\upsilon-\lambda\upsilon=du$, where $u=u^f$ is a harmonic function with boundary value $f$. Then
\begin{align*}
\Lambda f=du(\nu)=0-\lambda\upsilon(\nu), \quad \partial_\gamma f=du(\partial_\gamma)=\star\upsilon(\partial_\gamma)=-\upsilon(\nu).
\end{align*}
Hence, $\Lambda f=\lambda\partial_\gamma f$ and $Hf=\lambda f$. 

\smallskip

2) Each $a_j$ can be represented as $a_j=l_j-l'_j$, where the curves representing $l_j$ and $l'_j$ belong to $M$ and $\tau(M)$, respectively. The condition $\tau\circ a_j=-a_j$ implies $l'_j=\tau\circ l_j$, so $a_j=l_j-\tau\circ l_j$ while the linear independence and primitivity of $a_1,\dots,a_{2\mathfrak{g}}$ imply that $\{l_j\}_{j=1}^{2\mathfrak{g}}$ is a basis in $H_1(M, \partial M; {\mathbb Z})$. 

Introduce the Abelian differentials 
$$\omega_i:=\frac{1}{2}(1+i\star)\upsilon_i,$$
where $\upsilon_i$ are the elements of the dual (to $\{l_j\}_{j=1}^{2\mathfrak{g}}$) basis in $\mathscr{H}^1_0(M)$ extended harmonically onto $2M$. Then 
\begin{align*}
\int_{a_j}\omega_i=\frac{1}{2}\int_{l_j}(\upsilon_i-\tau^*\upsilon_i)+\frac{1}{2i}\int_{l_j}(\star\upsilon_i-\tau^*\star\upsilon_i)=\int_{l_j}\upsilon_i+0=\delta_{ij}
\end{align*}
due to the equalities $\tau^*\upsilon=-\upsilon$ and $\tau^*\star=-\star\tau^*$. 

Due to the Riemann bilinear relations $\Im\mathbb{B}_{ij}=\frac{i}{2}\int_{2M}\omega_i\wedge\overline{\omega_j}$, the imaginary part of the b-period matrix $\mathbb{B}$ of $2M$ does not depend on the choice of $b$-cycles complementary to $\{a_j\}_{j=1}^{2\mathfrak{g}}$. Thus, one can choose $b_j=\pm l_{j\pm\mathfrak{g}}$, where the sign obeys $j\pm\mathfrak{g}\in[1,2\mathfrak{g}]$. Then
$$2\Im\mathbb{B}_{ij}=2\Im\int_{b_j}\omega_i=\int_{\pm l_{j\pm\mathfrak{g}}}\star\upsilon_i=\pm\mathfrak{B}_{i,j\pm\mathfrak{g}}$$
and, in particular, ${\rm det}(2\Im\mathbb{B})={\rm det}\mathfrak{B}$. By this, Lemma \ref{Kolemma} is proved.\qed

\begin{rem}
If $\partial M$ has several connected components, then the approach of the present paper is ineffective for at least the following reason: the operator
 $\partial_\gamma^{-1}$ is defined on the orthogonal complement to functions that are constant on each connected component of $\partial M$, and, since ${\rm Ran}\Lambda=L_2(\partial M)\ominus\mathbb{C}$,  the Hilbert transform $H=\partial_\gamma^{-1}\Lambda$ is not defined. If one defines $\hat{H}=\Lambda^{-1}\partial_\gamma$, then applying the above method yields the equality ${\rm det}(\partial_\gamma)={\rm det}(\Lambda){\rm det}(\hat{H})$, which is trivial since ${\rm det}(\partial_\gamma)={\rm det}(\hat{H})=0$. Note that one cannot use the restriction of the equality $\partial_\gamma=\Lambda \hat{H}$ onto $L_2(\partial M)\ominus{\rm Ker}(\partial_\gamma)$ since this subspace is not invariant for $\Lambda$.
 
An expression for ${\rm det}\,\Lambda$  for bounded multi-connected plane domains in terms of the periods of their Schottky doubles is found in \cite{KokKor}  via a completely different method (which, in its turn, is not applicable to surfaces with connected boundary).   
 
\end{rem}

\end{document}